\documentclass[11pt,letterpaper,twoside]{article}
\usepackage{color,graphicx,natbib,lscape,here,geometry,setspace,multirow,enumitem}
\usepackage[dvipsnames]{xcolor}
\usepackage{graphicx}

\usepackage{authblk}
\usepackage{silence}
\usepackage{rotating}
\usepackage{multirow}
\usepackage{setspace}
\usepackage{algorithm2e}
\usepackage{enumitem}
\usepackage{tikz}
\usepackage[utf8]{inputenc}  
\usepackage{booktabs}
\usepackage{threeparttable}

\newcommand*{\thead}[1]{\multicolumn{1}{c}{\bfseries #1}}

\usepackage{amsmath}
\usepackage{amssymb}

\usepackage{newtxtext,newtxmath}
\newcommand*{\scfont}{\fontfamily{ptm}\selectfont}

\definecolor{nblue}{RGB}{108, 124, 89}
\usepackage[colorlinks=true,urlcolor=nblue,linkcolor=nblue,citecolor=nblue,backref=page]{hyperref}
\usepackage{bigstrut}

\usepackage{tabularx}
\usepackage{threeparttable}
\usepackage{dcolumn}

\usepackage{etoolbox} 
\makeatletter
\patchcmd{\BR@backref}{\newblock}{\newblock[}{}{}
\patchcmd{\BR@backref}{\par}{]\par}{}{}
\makeatother

\usepackage{pdflscape}
\usepackage{afterpage}
\usepackage{rotate}
\usepackage{longtable}
\usepackage{array}
\newcolumntype{C}[1]{>{\centering\arraybackslash}p{#1}}

\usepackage{comment}
\usepackage{mathtools}

\usepackage[hang,flushmargin]{footmisc} 
\usepackage[british]{babel}

\pdfoutput=1
 \usepackage{float}
 \restylefloat{table}

\usepackage[title,titletoc]{appendix}
\makeatletter

\renewenvironment{appendices}{%
    \begin{oldappendices}%
    \renewcommand{\thefigure}{\ifnum \c@section>\z@ \thesection.\fi\@arabic\c@figure}%
    \@addtoreset{figure}{section}%
    \renewcommand{\thetable}{\ifnum \c@section>\z@ \thesection.\fi\@arabic\c@table}%
    \@addtoreset{table}{section}}{%
    \end{oldappendices}%
}\makeatother

\usepackage{titlesec} 
\titleformat{\section}[block]{\large}{\thesection. }{0em}{\MakeUppercase} 
\titleformat{\subsection}[block]{\large}{\thesubsection. }{0em}{\itshape} 
\titleformat{\subsubsection}[block]{\large}{}{0em}{\itshape} 

\let\natbibcitet\citet
\renewcommand\citet{\bibpunct{(}{)}{,}{a}{,}{,}\natbibcitet}
\let\natbibcitep\citep
\renewcommand\citep{\bibpunct{(}{)}{;}{a}{,}{;}\natbibcitep}
\newcommand{\bi}{\begin{itemize}}
\newcommand{\ei}{\end{itemize}}
\newcommand{\be}{\begin{equation}}
\newcommand{\ee}{\end{equation}}
\defcitealias{ieo14}{IEO, 2014}

\usepackage{sectsty}
\allsectionsfont{\normalsize}

\makeatletter
\renewcommand\theequation{\thesection.\arabic{equation}}
\@addtoreset{equation}{section}
\makeatother
\makeatletter
\def\ubar#1{\underline{\sbox\tw@{$#1$}\dp\tw@\z@\box\tw@}}
\def\obar#1{\overline{\sbox\tw@{$#1$}\dp\tw@\z@\box\tw@}}
\makeatother

\usepackage{bm}
\usepackage[font=small]{caption}
\usepackage{subcaption}

\makeatletter\let\p@subfigure\thefigure\makeatother

\floatstyle{plaintop}
\restylefloat{table}

\usepackage{fancyhdr} 
\usepackage{cleveref}
\crefname{chapter}{Chapter}{Chapters}
\crefname{section}{Section}{Sections}
\crefname{subsection}{Section}{Sections}
\crefname{subsubsection}{Section}{Sections}
\crefname{figure}{Figure}{Figures}
\crefname{table}{Table}{Tables}
\crefname{equation}{Equation}{Equations}
\crefname{appendix}{Appendix}{Appendices}
\crefname{appendices}{Appendix}{Appendices}

\crefname{appsec}{Appendix}{Appendices}

\usepackage{catoptions}
\makeatletter

\def\Autoref#1{%
  \begingroup
  \edef\reserved@a{\cpttrimspaces{#1}}%
  \ifcsndefTF{r@#1}{%
    \xaftercsname{\expandafter\testreftype\@fourthoffive}
      {r@\reserved@a}.\\{#1}%
  }{%
    \ref{#1}%
  }%
  \endgroup
}
\def\testreftype#1.#2\\#3{%
  \ifcsndefTF{#1autorefname}{%
    \def\reserved@a##1##2\@nil{%
      \uppercase{\def\ref@name{##1}}%
      \csn@edef{#1autorefname}{\ref@name##2}%
      \autoref{#3}%
    }%
    \reserved@a#1\@nil
  }{%
    \autoref{#3}%
  }%
}
\makeatother

\newcolumntype{d}[1]{D{.}{.}{#1}}

\renewcommand\labelenumi{(\roman{enumi})}
\renewcommand\theenumi\labelenumi

\title{\huge{A Factor-Augmented Markov Switching (FAMS) Model}\thanks{Address: Department of Economics, Vienna University of Economics and Business. Welthandelsplatz 1, 1020 Vienna, Austria. E-mail: \texttt{\href{mailto:gregor.zens@wu.ac.at}{gregor.zens@wu.ac.at}} and \texttt{\href{maximilian.boeck@wu.ac.at}{maximilian.boeck@wu.ac.at}}. \textit{Date}: \today.}}
\author{\large{\uppercase{Gregor Zens} and \uppercase{Maximilian Böck} }\\
\vspace*{-0.5em}\normalsize{\textit{Vienna University of Economics and Business}}}
\date{}

\newcommand\given[1][]{\:#1\vert\:}


\def\equationautorefname~#1\null{%
  Eq.~(#1)\null
}
\def\equationautorefname~#1\null{
Eq.~(#1)\null
}

\onehalfspace
\begin{document}

\maketitle

\begin{abstract} 

This paper investigates the role of high-dimensional information sets in the context of Markov switching models with time varying transition probabilities. Markov switching models are commonly employed in empirical macroeconomic research and policy work. However, the information used to model the switching process is usually limited drastically to ensure stability of the model. Increasing the number of included variables to enlarge the information set might even result in decreasing precision of the model. Moreover, it is often not clear a priori which variables are actually relevant when it comes to informing the switching behavior. Building strongly on recent contributions in the field of factor analysis, we introduce a general type of Markov switching autoregressive models for non-linear time series analysis. Large numbers of time series are allowed to inform the switching process through a factor structure. This factor-augmented Markov switching (FAMS) model overcomes estimation issues that are likely to arise in previous assessments of the modeling framework. More accurate estimates of the switching behavior as well as improved model fit result. The performance of the FAMS model is illustrated in a simulated data example as well as in an US business cycle application.

\end{abstract}

\begin{keywords} Bayesian analysis, factor models, Markov switching, business cycles.\end{keywords}

\begin{codes} C11, C24, C38, E32, E37\end{codes}

\clearpage


\onehalfspacing
\renewcommand{\thepage}{\arabic{page}}

\section{Introduction}
\label{sec:introduction}

Since the late 1980s, Markov switching (MS) models have been widely employed by economic researchers to analyze various quantities of interest. Early studies analyze for instance stock market returns (\citealp{pagan1990alternative}) or asymetries over the business cycle (\citealp{hamilton1989}). It has been widely acknowledged that MS models are rather useful in capturing nonlinear time series characteristics as compared to more classical approaches such as standard ARMA models (\citealp{hamilton1994time}). As a result, MS models have gained in popularity and are commonly found in empirical macroeconomic research (\citealp{hamilton2010}). \\

Despite the popularity of the model family, the standard MS framework does not come without criticism. As pointed out by \citet{KAUFMANN201582}, a main critique is the assumption of exogenous transition probabilities. This assumption implies that there is no explicit interpretation of the process driving the switching behavior of the model. Moreover, the most commonly applied prior framework a priori favors the model switching to the states that are visited most often. This approach leads to a model that neglects further information as for instance macroeconomic conditions that are readily available to the researcher.\\ 

This criticism has lead to a rather popular extension of the standard MS model that uses time-varying transition probabilities arising from a prior structure taking the form of a (multinomial) logit or probit specification (\citealp{filardo1994business}; \citealp{Meligkotsidou2011}; \citealp{KAUFMANN201582}).  This modified prior setup effectively enables relevant exogenous variables to inform the switching behavior of the model. 
\citet{KAUFMANN201582} discusses Bayesian estimation and models a Philipps curve in a MS framework.\\

While in general this method is a viable and useful extension of the standard MS framework, it comes with some unpleasant downsides, especially when dealing in data rich environments like macroeconomics or finance. The (multinomial) probit or logit regression that forms the prior distribution of the transition probabilities is known to perform rather poorly when the set of predictors becomes too large as discussed in \citet{zahid2013multinomial}, \citet{ranganathan2017common} and \citet{dejong2019}. This often destabilizes the modeling framework and makes the effort to model regime switching using exogenous regressors cumbersome. In addition, it is often not clear a priori which variables are relevant to include when modeling switching behavior in a MS setting. In principal, this problem can be overcome by shrinkage priors on the coefficients of the logit/probit regression as demonstrated in \citet{Zens2019} for the related family of mixture-of-experts models. However, although variable selection might work well for a medium sized set of predictors, it is likely to become problematic when the number of regressors becomes too large compared to the available information in the data. Even more importantly, a variable selection approach does not resolve multicollinearity issues that commonly arise in macroeconomics and finance, where various time series describe very similar underlying processes of an economy. For instance, quarterly gross domestic product and quarterly industrial production are likely to show a large amount of comovement. Variable selection is prone to be not well behaved in such cases (\citealp{george2010dilution}). Thus, the researcher is required to come up with a pre-selection of relevant variables. This process is usually rather arbitrary and likely to severely reduce the information set available to inform the switching behavior of the model.\\

As a result, most articles dealing with MS models in economics and finance restrict the number of variables that inform the switching process to a relatively small number.\footnote{For instance, \citet{KAUFMANN201582} uses one exogenous variable and \citet{Meligkotsidou2011} uses six variables to inform the switching process.} As pointed out by \citet{bernanke2005favar}, a small number of variables will most probably not be able to span the information set consisting of hundreds of time series available to the researcher, to central banks and to financial market participants. This leads to two potential problems when employing these ''restricted'' data sets. The first problem relates to policy analysis and forecasting exercises. It is well known that despite good in-sample predictions, out of sample forecasts of MS models regularly fail to consistently beat simple benchmark processes (\citealp{engel1994can}; \citealp{boot2017macroeconomic}). In these cases, the models might be missing important information due to restrictions in the estimation process. However, as shown by \citet{stock2002forecasting}, the forecasting ability of models using factor augmentation increases significantly as compared to models using ''restricted'' information sets. Second, the effect of certain variables of interest to researchers and policy makers cannot be evaluated in the ''restricted'' framework as they are not explicitly included in the model.\\

To overcome above mentioned issues, we offer a modeling framework that combines standard MS models with recent developments in factor modeling. In spirit, this article is closely related to the FAVAR model developed in \citet{bernanke2005favar} and builds on the sparse factor model outlined in \citet{KASTNER2018}. The factor augmented Markov switching model (FAMS) that we derive in this article effectively summarizes large amounts of information about the economy and financial markets in a small number of estimated factors. Augmenting the model to inform the switching process in a MS framework provides a rather appealing solution to computational and statistical problems arising when using a large number of variables in this modeling framework.\\

This article summarizes the key points of the statistical framework necessary to estimate FAMS models in a fully Bayesian setup through Gibbs sampling. We demonstrate the ability of this model using autoregressive processes with possibly switching means and switching variances. Applying the model to an artificial data set, we find that the large information set included in FAMS models results in improved in-sample fit and more precise state allocations as compared to standard MS models making use of the full data set or random subsets of the data. The real world abilities of FAMS models are demonstrated through an application, where we aim to identify US business cycles through informing the switching process with a vast amount of macroeconomic time series.\\

Our contribution is thus twofold. The FAMS model enables it to conveniently estimate MS models where the switching process is influenced by a lower dimensional representation of a large number of possibly collinear candidate variables. Moreover, a shrinkage prior is imposed to conduct variable selection in cases where some factors might be irrelevant for the regime switching behavior of the model.\\

The remainder of this article is organized as follows. In Section 2 we formulate the general modeling framework for the FAMS model. Section 3 discusses Bayesian estimation of the model using MCMC methods. Section 4 presents the benefits of the FAMS model via a simulation study. In Section 5, a high-dimensional macroeconomic data set is used in an application to estimate business cycles in the United States. A brief conclusion is provided in Section 6.\\

\section{Factor-Augmented Markov switching Model}
\label{sec:framework}
\subsection{Markov switching AR(p) Model with Time Varying Transition Probabilities}
\label{sec:ms}

We consider a general Markov switching autoregressive model (MS-AR) of order $p$ with time-varying transition probabilities. This model has been thoroughly discussed in \citet[Ch. 4; Ch. 9]{kim1999state} and \citet[Ch. 12]{fruhwirth2006finite}. Let $\{y_{t}\}_{t=1}^T$ be an observed time series arising from the data-generating process (DGP)
\begin{equation}
\label{eq:ms-ar}
		y_{t} = \mu_{S_{t}} + \sum^p_{j=1} \phi_{j,S_{t}} y_{t-j} + \epsilon_{t}, \qquad \epsilon_{t} \sim N(0,\sigma^2_{S_{t}}),
\end{equation}

where $\{S_{t}\}_{t=1}^T$ is assumed to be a hidden discrete-time state process with finite state space $\mathcal{H} = \{1,\hdots,H\}$. In the most general setup, the intercept parameter $\mu_{S_{t}}$, the persistence parameters $\phi_{j,S_{t}}$ and the error variance parameter $\sigma^2_{S_{t}}$ are state-dependent. First, we collect all our parameters as $\theta = \{\theta_h | \theta_h = (\mu_h,\phi_{1,h},\hdots,\phi_{p,h},\sigma_h^2)^\top, h = 1,\hdots,H\}$. Thus, it is straightforward to see that if the process is in state $k \in \mathcal{H}$ in time period $t$, parameter values change accordingly:

\begin{equation}
\label{eq:state-dependence}
    \theta_k = (\mu_k, \phi_{1,k},\hdots,\phi_{p,k},\sigma_k^2) \quad \text{if} \quad S_t = k.
\end{equation}

The peculiarity of Markov switching models is that their switching is determined completely by a stochastic process. The state indicator vector $S$ can thus be seen as an irreducible, aperiodic Markov chain describing a sequence of events where the probability of each state only depends on the state attained in previous rounds. We assume a first order Markov process, where the process has only a memory of one period:

\begin{equation}
\label{eq:markovprocess}
    Pr\lbrack S_t=j | S_{t-1} = h_{t-1}, S_{t-2} = h_{t-2}, \hdots, S_{1} = h_1 \rbrack = Pr\lbrack S_t = j | S_{t-1} = k\rbrack.
\end{equation}

Therefore, the transitions between two states are described by probabilities that are summarized in a transition matrix $\Xi_t$ of dimensions $H \times H$. In this transition matrix, each element $\xi_{jk,t}$ sufficiently describes the stochastic behaviour, i.e. the transition from state $k$ to state $j$. Note that the matrix $\Xi_t$ is row-standardized. Furthermore, the transition matrix $\Xi_t$ features a time index $t$ since we assume that the switching process is not constant over time. It is modeled as dependent on a set of $r$ exogenous factors captured in $\{Z_t\}_{t=1}^T$. These variables are not entering the model directly in the autoregressive process, but are rather expected to drive the transition dynamics indirectly. Therefore, they are responsible for altering the state allocation and thus the estimation of our state-dependent parameters in $\theta$. Moreover, a delay parameter $d$ is introduced for convenience in cases where the exogenous factors $Z_t$ are used as leading indicators. The estimation of $Z_t$ is covered in section \autoref{sec:dfm}.\\

The multinomial logit link is used to model the influence of the factors $Z_t$ on the transition probabilities $\xi_{jk,t}$ in a way such that
\begin{equation}
\label{eq:mnl}
Pr\lbrack S_{t} = j | S_{t-1} = k, Z_{t-d}, \gamma\rbrack = \xi_{jk,t} = \frac{\text{exp}(\gamma_{jk} + \beta_{jk} Z_{t-d})}{\sum^H_{l=1}\text{exp}(\gamma_{jl} + \beta_{jl} Z_{t-d})},
\end{equation}
where $j,k = 1,\hdots,H$ and $\gamma_{jk}$ and $\beta_{jk}$ are the coefficients of the intercepts and the factors, respectively. Moreover, we follow \citet{KAUFMANN201582} and specify $Z_{t} = \tilde{Z}_{t}-\bar{Z}_i$ in a centered way for two reasons. Besides the fact that it defines the mean $\bar{Z}_i$ as an arbitrary threshold level, more importantly, the time-invariant part of the transition probabilities $\gamma_{jk}$ does then not depend on the scale of $\tilde{Z}_{t}$. Otherwise, this would have to be taken into account when choosing a prior distribution for the parameter.\\

For identification purposes, it is necessary to define a baseline state $h_0 \in \mathcal{H}$ in which the parameters are assumed to be zero, i.e. $(\gamma_{jh_0},\beta_{jh_0}) = 0$. This yields
\begin{equation*}
Pr\lbrack S_{t} = j \given S_{t-1} = h_0, Z_{t-d},\gamma\rbrack = \frac{1}{1 + \sum_{l \in \mathcal{H}_{-h_0}}\text{exp}(\gamma_{jl} + \beta_{jl} Z_{t-d})},
\end{equation*}
where $\mathcal{H}_{-h_0}$ denotes all states but the reference state $h_0$. \\

In order to improve the efficiency of the estimates, we follow \cite{AMISANO2013118} in imposing the restriction of common slope coefficients across states, $\beta_{jk} = \beta_{k}$. This translates into the assumption that the effect of $Z_{t}$ differs only by the state attained in the previous time period.
For notational convenience, we collect the multinomial logit coefficients in the vectors $\gamma = \{\gamma_h \given \gamma_h = (\gamma_{1h},\hdots,\gamma_{Hh})^\top, h = 1,\hdots,H\}$ and $\beta = \{\beta_h, h = 1,\hdots,H\}$ and gather all parameters in $\vartheta = \{\theta, \gamma, \beta\}$.

\subsection{Factor Model}
\label{sec:dfm}

In the above mentioned modeling framework, it is possible to proceed by estimating the model using a given subset of the full set of exogenous covariates $X_t$ to inform the switching process through the prior specification. However, there is a variety of scenarios where additional relevant information may be available that is not incorporated in this subset. Suppose we can find a set of latent factors $Z_t$ that condenses a large part of the available information into a small number of time series. These factors might represent more diffuse concepts such as ''political instability'' or ''credit market sentiments''. These concepts are generally difficult to capture in one or two time series. The task of finding $Z_t$ given $X_t$ can be dealt with using factor analysis frameworks (see \citealp{lopes2004bayesian} or \citealp{lopes2014modern} for a comprehensive overview). The setup used in this article is closely following \citet{KASTNER2018} who proposes a time varying sparse factor model. It can be summarized as

\begin{equation}
\begin{split}
        X_t|\Lambda, Z_t, U_t &\sim N(\Lambda Z_t, U_t)\\
Z_t|V_t &\sim N(0, V_t)
\end{split}
\end{equation}

where $Z_t$ is the set of latent factors we want to recover for use in the MS model, $\Lambda$ is a $m \times r$ factor loadings matrix and $U_t$ and $V_t$ are diagonal matrices discussed in more detail below. The $m$ scaled and centered time series $X_t$ = $(X_{1t},\ldots,X_{mt})$ describing the economy are assumed to follow a conditional Gaussian distribution, i.e.

\begin{equation}
    X_t|\Sigma_t \sim N_m(0,\Sigma_t),
\end{equation}
where $N_m(\cdot)$ denotes a $m$-dimensional normal distribution. To achieve dimensionality reduction, the $m \times m$ covariance matrix $\Sigma_t$ is assumed to decompose into a $m \times r$ factor loadings matrix $\Lambda$ as well as two diagonal matrices $V_t$ $(r \times r)$ and $U_t$ $(m \times m)$ as follows:

\begin{equation}
    \Sigma_t = \Lambda V_t \Lambda' + U_t.
\end{equation}

Following \citet{KASTNER2018}, $\Lambda$ is assumed to be constant over time. The factor variances in $V_t$ and the idiosyncratic variances in $U_t$ are evolving over time through stochastic volatility models (\citealp{kastner2014ancillarity}). Specifically, let $U_t = \text{diag}(\text{exp}(g_{1t}), \ldots, \text{exp}(g_{mt}))$ and $V_t = \text{diag}(\text{exp}(h_{1t}), \ldots, \text{exp}(h_{rt}))$. Then the log variances are modeled through autoregressive processes of the form

\begin{equation}
    g_{it} \sim N(\mu_{g,i} + \phi_{g,i}(g_{it-1} - \mu_{g,i}), \sigma_{g,i}^2), \qquad i = 1, \ldots, m
\end{equation}

and

\begin{equation}
   h_{jt} \sim N(\phi_{h,j}h_{jt-1}, \sigma_{h,j})^2, \qquad j = 1, \ldots, r.
\end{equation}

That is, the idiosyncratic variances follow a centered AR(1) process and the factor variances follow an autoregressive process with zero mean to identify the scaling of the factors. For further details on the factor model described here, refer to \citet{KASTNER2018}. For further topics in factor modeling including estimation and identification issues, see for instance \citet{lopes2014modern} or \citet{kaufmann2019bayesian}.

\section{Bayesian estimation}
\label{sec:est}
\subsection{Prior setup}
\label{sec:prior}

Bayesian estimation requires the specification of adequate prior distributions. This section gives an overview of prior distributions both in the Markov switching model (\autoref{sec:ms}) and the factor model (\autoref{sec:dfm}) that form the proposed model setup.\\

For the Markov switching AR(p) model, independent prior distributions for all parameters $\pi(\vartheta) = \pi(\alpha)\pi(\phi)\pi(\sigma^2)\pi(\gamma)\pi(\beta)$ are assumed. Since the specification in \autoref{eq:ms-ar} is piece-wise linear, we specify a normal prior distribution for the mean and persistence parameters and an inverse Gamma distribution for the variance parameters

\begin{equation}
\label{eq:prior:ms}
\begin{aligned}
    \mu &\sim \prod^H_{h=1} \pi(\mu_h) = \prod^H_{h=1} N(m_0, M_0),\\
    \phi_j &\sim \prod^H_{h=1} \pi(\phi_{j,h}) = \prod^H_{h=1} N(r_0,R_0), \quad \forall j=1,\hdots,p,\\
    \sigma^2 &\sim \prod^H_{h=1} \pi(\sigma^2_h) = \prod^H_{h=1} IG(c_0,d_0).
\end{aligned}
\end{equation}

Regarding the hyperparameters, we choose $m_0 = r_0 = 0$, $M_0 = 10$ and $R_0 = 4$ to stay rather uninformative. In setting $c_0=d_0=1$, we use an informative prior distribution on the variance parameters in order to regularize the variances sufficiently away from zero to circumvent singularities.

In the multinomial logit, seperate prior distributions for the intercepts $\gamma$ and the coefficients $\beta$ are specified. For the intercepts, we assume

\begin{equation}
\label{eq:prior:mnl:intercept}
    \gamma \sim \prod^H_{h \in \mathcal{H}_{\text{-}h_0}} \pi(\gamma_h) = \prod^H_{h \in \mathcal{H}_{\text{-}h_0}} N(g_{0,h},G_0), \\
\end{equation}
where $g_{0,h}=0$ and $G_0=4$. 

Generally speaking, some factors might be more prone to capture information relevant to the switching process, whereas other factors might mainly introduce noise to the model. To overcome these issues, a normal gamma shrinkage prior (\citealp{polson2010}) is chosen for the coefficients $\beta$ of the exogenous factors. Therefore, the prior on each element for each state of the coefficient vector can be written as

\begin{equation}
\label{eq:prior:mnl:coefs}
    \beta_{i,h}\given \psi_{i,h} \sim N(0, \frac{2}{\lambda^2_{\psi,h}}\psi_{i,h}),\quad \psi_{i,h} \sim G(\omega_\psi,\omega_\psi), \quad \forall h \in \mathcal{H}_{\text{-}h_0}, \quad \forall i=1,\hdots,r.
\end{equation}

Now we turn to the specification of prior distributions for the factor model. For the elements of the factor loadings matrix, again a normal gamma prior is employed to achieve sparsity:
\begin{equation}
\label{eq:prior:dfm}
    \Lambda_{i,j} \given \tau_{i,j} \sim N(0, \frac{2}{\lambda^2_{\tau,i}}\tau_{i,j}), \quad \tau_{i,j} \sim G(\omega_\tau,\omega_\tau), \quad \forall i=1,\hdots,m \quad \forall j=1,\hdots,r
\end{equation}
where the global shrinkage parameter $\lambda^2_{\tau,i}$ is defined per row of the factor loadings matrix. Thus, this setup implies that each time series has a high a priori probability of not loading on any factor. Similarly, the global shrinkage parameter $\lambda^2_{\psi,h}$ is defined per equation of the multinomial logistic prior to allow varying levels of shrinkage across states in the MS model. In choosing prior distributions on the stochastic volatility parameters, we strongly follow \citet{KASTNER2018}. Since both the stochastic volatility process of the factor variances and the idiosyncratic variances have own persistence and variance parameters, a subscript $j \in \{g,h\}$ indicates to which group of processes they belong. The priors can then be summarized as
\begin{equation}
\begin{aligned}
    \mu_g &\sim N(0, 100), \\
    \frac{\phi_j + 1}{2} &\sim B(b_0,b_1), \\
    \sigma_j^2 &\sim G(1/2, 1/(2B_{\sigma})),
\end{aligned}
\end{equation}

where the function of persistence parameters $\phi_j$ follows a Beta distribution to ensure stationarity.\\

Finally, it is useful to impose hyperpriors on the hyperparameters in \autoref{eq:prior:mnl:coefs} and \autoref{eq:prior:dfm}. In general, normal gamma type shrinkage priors \citep{polson2010,griffin2010} enable implicit variable selection in a sophisticated way by specifying a continous prior distribution with a rather large probability mass on zero and fat tails. The degree of shrinkage is influenced by two parameters: a global shrinkage parameter $\lambda_j$ ($j \in \{\psi,\tau\})$ and a local shrinkage parameter $\psi$ or $\tau$. As seen in \autoref{eq:prior:mnl:coefs} and \autoref{eq:prior:dfm}, an additional layer of prior distributions on the local shrinkage parameters is introduced. Both $\psi$ and $\tau$ are assumed to follow a Gamma distribution with shape and scale hyperparameters $\omega_j$ ($j \in \{\psi, \tau\})$. Note that by deviating from specifying the shape and scale parameters to be equal, it is not possible to identify either the global shrinkage parameter or the parameters of the prior for the local shrinkage component. A choice of $\omega_j = 1$ leads to the Bayesian LASSO (\citealp{park2008bayesian}). Smaller values are associated with a more severe penalty function and thus impose stronger shrinkage. To complete the prior setup, a Gamma prior is specified on the global shrinkage component $\lambda_j^2$, $j \in \{\psi,\tau\}$:

\begin{equation}
\label{eq:prior:ng}
\lambda_j^2 \sim G(c_{j0},c_{j1}),
\end{equation}

where the hyperparameters for the prior on $\beta$ are set to $c_{\psi0}=c_{\psi1}=0.01$.  This implies heavy shrinkage on $\beta$. For reference, see for instance \citet{bitto2019achieving}. The hyperparameter values in the factor model are chosen in accordance with the standard values proposed by \citet{kastnerfacstochvol}. They imply less heavy shrinkage on the elements of the factor loadings matrix $\Lambda$.

\subsection{Posterior Simulation Using MCMC}
\label{sec:post}

Estimation of the model is carried out via MCMC sampling using a variety of data augmentation techniques. Bayesian estimation and inference with a Gibbs sampling algorithm requires the combination of the likelihood and the proposed priors to produce conditional posterior distributions for all parameters. The employed algorithms closely follow the algorithms described in \citet[Ch. 11]{fruhwirth2006finite} and \citet[Ch. 9]{kim1999state}.  Since the FAMS model is a variant of a state space model, the data augmentation techniques introduced by \citet{carter1994gibbs} and \citet{fruhwirth1994data} are viable candidates for performing estimation. Sequentially sampling from the conditional posterior distributions after convergence of the algorithm is achieved then allows posterior parameter inference to take place. Generally speaking, the sampling scheme employed in this paper is designed to iterate over the following three steps:

{\begin{enumerate}
  \setlength{\itemsep}{0pt}
  \setlength{\parskip}{0pt}
	\item \textit{Classification}. Conditional on the estimated parameters, the filtering approach proposed by \citet{hamilton1989}, a forward-filtering backward-sampling scheme, is employed to classify each observation $t$ into one of the $H$ states by sampling the state indicator from the posterior distribution $p(S \given y, Z, \vartheta)$.
	\item \textit{Estimation}. Conditional on the state indicators vector $S$, the regime-specific parameters $(\theta_1,\hdots,\theta_H)$ and $(\gamma,\beta)$ are conditionally independent. Therefore, we rely on standard Bayesian techniques for linear models to draw from the posterior $p(\theta \given S, y)$ and use the partial dRUM approach described in \citet{fruhwirth2010data} to simulate the multinomial logit coefficients $p(\gamma,\beta \given S, Z)$.
	\item \label{item:post:identification} \textit{Identification}. Label switching is a widely known issue one has to consider when working with mixture and Markov switching models. Since this is a non-trivial issue, a discussion is provided in Section \autoref{sec:est:ident}.
\end{enumerate}}

For estimating the factor stochastic volatility related quantities, we use the \texttt{R}-package \texttt{factorstochvol} provided by \citet{kastnerfacstochvol}.\footnote{In this preliminary version of the paper we choose to implement the FAMS model in a two step procedure. In a first step, we simulate from the posterior distributions of the factors. The resulting posterior means are then used as explanatory variables that inform the switching process in the MS framework. A fully Bayesian approach that includes the uncertainty around the factor posterior means will thus result in slightly higher uncertainty around other model estimates and will be included in future versions of the paper.} This allows us to make use of the efficient implementation using interweaving techniques. For ease of reference, estimation of the multinomial logistic prior specification using the partial dRUM sampler is discussed in detail in \autoref{sec:mnl_estimation}. Posterior simulation resulting from a normal gamma prior setup is briefly discussed in \autoref{ngprior}. The remaining simulation steps are standard and thus not discussed in detail. 

\subsection{Identification of the MS model}
\label{sec:est:ident}

Parameter estimation in the family of mixture and Markvo switching models can suffer from various difficulties, especially in a Bayesian framework. Label switching is a common issue in mixture modeling (\citealp{hurn2003estimating}; \citealp{jasra2005markov}). It is the result of the likelihood function being invariant to relabeling the components due to multimodality as discussed in \citet{redner1984mixture}. This can lead to problems as label switching during MCMC sampling might result in possibly distorted, multimodal posterior distributions that are in general difficult to summarize. Deriving posterior means or other point estimates based on these posteriors then becomes inappropriate (\citealp{stephens2000dealing}). The same rationale holds true for Markov switching models.\\ 

Early references for relabeling algorithms include \citet{celeux1996stochastic}. However, this algorithm requires known true parameter values, which renders it not very useful in real data applications. \citet{stephens2000bayesian} suggests an algorithm that relabels the draws in a way such that the (marginal) parameter posterior distributions are as unimodal as possible. \citet{stephens2000dealing} provides a literature review as well as a decision theoretic framework to deal with label switching.\\

In the simulation studies and application presented below, we make use of identifying restrictions to identify the sampler (see for instance \citealp{lenk2000bayesian}).\footnote{Knowing that this is not ideal (\citealp{fruhwirth2004estimating}), a random permutation sampler in combination with a post-processing procedure to relabel the draws will be implemented in future versions of this paper.}  This completes the simulation setup and the description of the employed estimation techniques of the FAMS model. The following sections discuss the model in the context of artificial data as well as an application to US business cycle estimation.

\section{Simulation Study}
\label{sec:simstudy}

To test the performance of the proposed modeling framework, synthetic data sets are generated using a recursive procedure. The factor model outlined in \autoref{sec:dfm} is used to simulate data under the assumption that the true DGP is driven by the factors.\footnote{We proceed like this for two reasons. The first argument is a theoretical one, where we argue that it is the factors which convey the information on whether transition probabilities should change or not. Second, from an econometric point of view generating probabilities from multinomial logit models using about 200 time series is neither advisable nor applicable. The stability of the exercise depends strongly on the amplitude of the effects and it is extremely tedious to generate a setup where a simulation study can be conducted in a controlled setup.} Hence, we proceed in the following manner: In a first step, we generate $r=3$ AR(1) processes

\begin{equation}
    f_{i,t} = 0.7 f_{i,t-1} + \eta_{i,t}, \qquad \eta_t \sim N(0,1). \quad i=1,\hdots,r,
\end{equation}

centered around zero. From these, 200 time series are generated using the DGP outlined in section \autoref{sec:dfm}. The AR coefficients $\phi_j$ $(j \in \{g,h\})$ of the log variances are simulated from U(-0.8,0.8) and the means $\mu_g$ are assumed to follow a normal distribution with N(0.2,0.2). The variances $\sigma^2_j$ $(j \in \{g,h\})$ are simulated from |N(0.2,0.2)|.\\

After that, three factors are estimated from the generated time series. Estimation is based on 80.000 draws, where the first 40.000 draws are discarded as burn-in. The true factors $F_t = (f_{1,t},f_{2,t},f_{3,t})^T$ are used to generate transition probabilities
\begin{equation}
    \xi_{jk,t} = \frac{\text{exp}(\gamma_{jk} + \beta_k F_t)}{\sum^H_{l=1} \text{exp}(\gamma_{jl} + \beta_{l}F_t)},
\end{equation}
which are then employed to draw the states $S_t$ of the process. Conditional on knowing the states, the final time series $y_t$ can be generated using the process

\begin{equation}
\label{eq:sim:ms-ar}
		y_{t} = \mu_{S_{t}} + \phi y_{t-1} + \epsilon_{t} \qquad \epsilon_{t} \sim N(0,\sigma^2_{S_{t}}).
\end{equation}

In this setup, $H=2$ and thus our finite state space is $\mathcal{H} = \{1,2\}$. We choose $\gamma_{j1} = (1.5,-1.5)^T$ and set $\gamma_{j2} = (0,0)^T$ for identification. Furthermore, $\beta_1 = (-1.2,1.1,0.9)^T$. Again, the second column of $\beta$ is set to zero due to identification reasons. The parameters in the AR model are set to $\mu = (-0.25,0.25)^T$, $\phi=0.55$ and $\sigma^2 = (0.1,0.05)^T$. Note that this setup allows to identify the sampler using identifying restrictions. Furthermore, the parameters are carefully chosen in a way such that a lot of time variation in the transition probabilities results and that the data is comparable to real world examples in economics and finance. This setting is used to generate $N=100$ different datasets, each with $T=250$.\\

Five different modeling approaches are implemented and compared. The first model imposes constant transition probabilities by using no external covariates in the MNL part of the MS model. This ''intercept only'' model serves as baseline model in what follows below. In the second model, the full information set of 200 time series is used to model the transition probabilities.\footnote{A Gaussian prior with zero mean and a variance of 4 is assumed on $\beta$ for the models with no shrinkage prior.} The third model uses all 200 time series as well, however, has a shrinkage prior placed on $\beta$ to induce sparsity. The fourth model corresponds to the FAMS model without normal gamma prior. Finally, the fifth model is the FAMS model including a normal gamma prior. To gain even more insight, all models are estimated with varying degrees of shrinkage induced by the normal gamma prior. For the sake of brevity, we fix the number of factors in the estimation to the true number of factors.\\

For each run of the MCMC sampler, $M=50,000$ draws are kept after a burn-in phase of $50,000$ draws. Two measures are employed to compare the performance of the models. The first one is the average median root mean squared error (RMSE) of the in-sample fit through
\begin{equation}
    RMSE_{\hat{y}} = \frac{1}{N} \sum^N_{n=1} q_{50}\Bigg(\sqrt{\frac{1}{T} \sum^T_{t=1} (y - \hat{y}^{(m)})^2}\Bigg).
\end{equation}
The second criterion is a measure of the misclassification rate of the states through
\begin{equation}
    MCR_{\hat{S}} = \frac{1}{N} \sum^N_{n=1} q_{50}\Bigg(\bigg(1-\frac{1}{T} \sum_{t=1}^T \mathcal{I}(\hat{S}_t = S_t)\bigg)\Bigg),
\end{equation}
where $\mathcal{I}(\cdot)$ denotes the indicator function. In both cases, $q_{50}$ denotes the posterior median.\\

\autoref{tab:SimStudy_results} reports the average in-sample median RMSE and MCR across 100 runs relative to the results of the baseline model. Results are presented for both modeling approaches and varying degrees of shrinkage as indicated by various values of $\omega_{\psi}$. It is noticeable that the FAMS model performs considerably better than the baseline model. Moreover, the FAMS model performs better than the full data set in most cases. A few interesting patterns are worth mentioning. Unsurprisingly, the full information set without a shrinkage prior performs even worse than the basline model (bear in mind that the full information set consists of 200 time series). Applying shrinkage therefore definitely makes sense and improves in-sample fit and classification. Furthermore, the degree of shrinkage matters. While the performance of $\theta_\psi=0.6$ and $\theta_\psi=1$ is relatively similar, inducing too much shrinkage (say, $\theta_\psi < 0.5$) seems not like an advisable strategy. In these cases, the prior tends to shrink too much information to zero. Hence, the performance of the full information set model approaches the baseline model as $\beta$ gets close to $0$. In our experience, the coefficients in a multinomial logistic prior in a MS context are rather sensitive and in general difficult to estimate precisely. Often, this leads to the normal gamma setup applying too much shrinkage. This furthermore underlines the benefits of using estimated factors in a MS context. It drastically reduces multicollinearity issues and maximizes the amount of information contained in a given time series used to estimate the multinomial logistic regression. Both channels may counteract problematic and unintended shrinkage effects.\\

\begin{table*}[t]
\centering
\begin{threeparttable}
        \caption{Simulation Study Results}
\label{tab:SimStudy_results}
\setlength\tabcolsep{15pt}
\renewcommand\arraystretch{1.3}
  \begin{tabular}{cccc}
    \toprule
   & & \thead{Average RMSE} &  \thead{Average MCR} \\ 
   \cmidrule(lr{1em}){2-4}
   \multirow{5}{*}{Full information set} & $\text{NG off}$ & $1.043$ & $1.011$\\ 
   & $\omega_\psi=0.2$ & $0.960$ & $0.640$ \\
   & $\omega_\psi=0.6$ & $0.933$ & $0.599$ \\
   & $\omega_\psi=1.0$ & $0.934$ & $0.601$ \\
   \cmidrule(lr{1em}){2-4}
   \multirow{5}{*}{FAMS} & $\text{NG off}$ & \thead{0.908} & \thead{0.525} \\ 
   & $\omega_\psi=0.2$ & $0.917$ & $0.608$ \\
   & $\omega_\psi=0.6$ & $0.912$ & $0.579$ \\
   & $\omega_\psi=1.0$ & $0.911$ & $0.575$ \\
\toprule
\end{tabular}
\end{threeparttable}
\end{table*}

In an additional exercise, we tried estimating the model using random subsamples from the full information set. This is supposed to roughly emulate theory-guided a priori variable selection. Each subsample consists of ten time series. Comparing the results of the best performing subsample with the FAMS model, two results are noticable: In-sample fit as measured by RMSEs become comparable to the FAMS model. Interestingly, model performance in terms of MCRs is rather bad. Here, further investigation is indicated.\\

Focusing on the FAMS model, the best performance is achieved in the setup without shrinkage prior. Again, we argue that the prior might shrink necessary information toward zero. However, more experimentation and simulation studies will be necessary to systematically investigate this issue. Apart from this observation, it has to be noted that the differences between various FAMS setups are very modest in general. We suspect that performance will strongly depend on the application at hand. Moreover, since the focus here lies not on variable selection, but on the benefits of projecting large information sets into lower dimensions, we conclude that the FAMS model does a pretty decent job.

\section{Application: Business Cycle Analysis using US Data}
\label{sec:app1}

Since the seminal paper of \citet{hamilton1989}, Markov switching models have been used to detect and explore the nature of business cycles. An excellent overview is provided by \citet{hamilton2016macroeconomic}. It is nowadays widely acknowledged that the nature of macroeconomic behaviour is usually not well captured by linear models. 

From a theoretical point of view, there is a lively debate on why and how regime changes take place. One school of thought posits that the alternation of expansions and recessions results from different paces of technological innovations, which are assumed to be exogenous to the economy. Another paradigm starts by assuming that the market economy is inherently instable and produces business cycles endogenously. A recent theoretical contribution by \citet{matsuyama2013good} thus endogenizes credit markets and shows how an economy itself can drive recurrent booms and busts, with each bust sowing the seeds of the next boom.

\subsection{Data \& Model}

For the basic model of the macroeconomy of United States, we obtain publicly available data from the Federal Reserve Bank of St. Louis \citep[see][]{doi:10.1080/07350015.2015.1086655}. This very well known data set has been used in a variety of macroeconomic studies, see for instance \citet{ludvigson2015uncertainty} or \citet{stock2016dynamic}. It covers a variety of macroeconomic series capturing eight core areas of the US economy such as real activity and financial market. Over 200 time series in this data set span an information set that is too large to be employed in standard MS models. Hence, the FRED database is a proper candidate to demonstrate the benefits of the FAMS model.\\

As the goal is to model business cycles, the quarterly log differenced time series of industrial production is modeled as uncentered AR(4) process with switching intercept:

\begin{equation}
\label{eq:spec-bc}
    y_t = \mu_{S_t} + \sum^4_{j=1} \phi_j y_{t-j} + \epsilon_t, \quad \epsilon_t \sim N(0, \sigma^2).
\end{equation}

\begin{figure}[t]
\begin{subfigure}{\textwidth}
  \centering
  \includegraphics[width=0.8\linewidth]{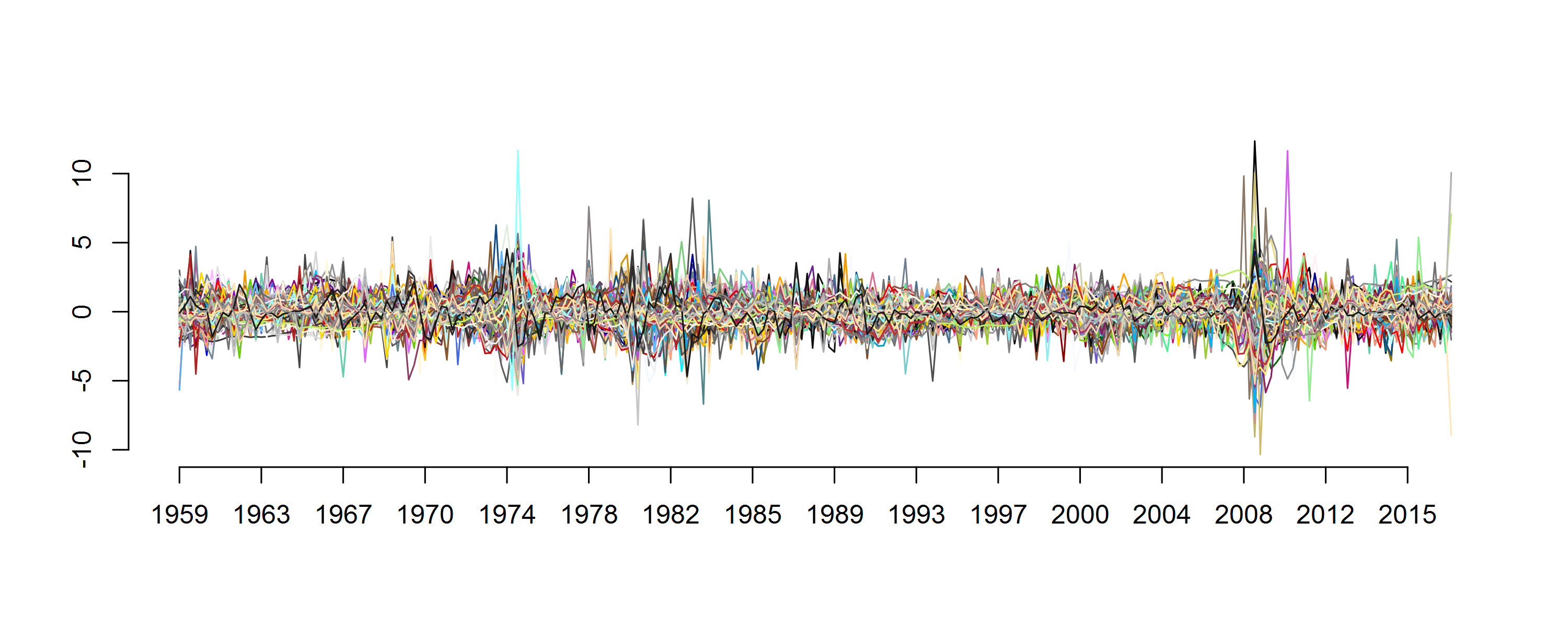}
  \caption{Full information set.}
  \label{fig:infoset}
\end{subfigure}\\
\begin{subfigure}{\textwidth}
  \centering
  \includegraphics[width=0.8\linewidth]{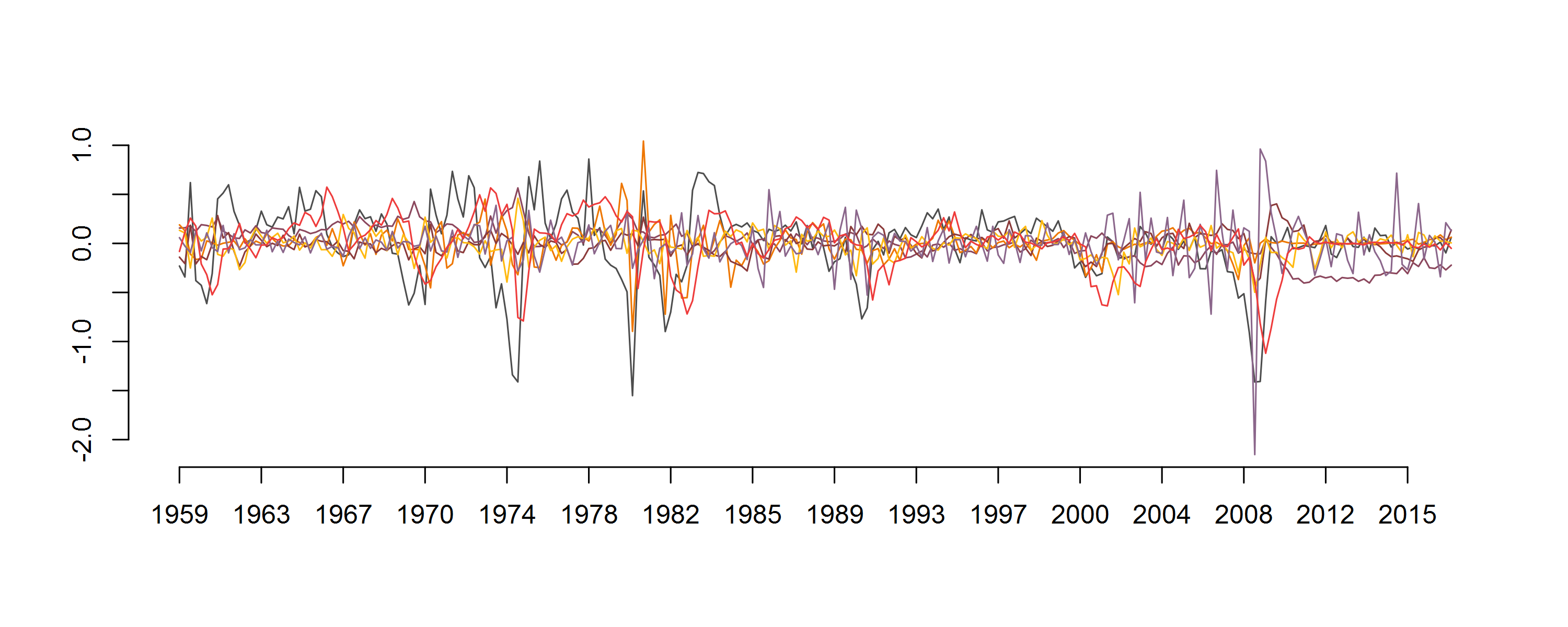}
  \caption{Posterior mean of extracted factors.}
  \label{fig:factors}
\end{subfigure}
\caption{Input and output from factor model.}
\label{fig:infoset_factors}
\end{figure}

This approach closely follows the seminal paper by \citet{hamilton1989}, who however uses a centered MS-AR model. We deviate from the centered parameterization, because in a Markov switching context the uncentered parametrization does not induce immediate mean level shifts. Instead, the mean level approaches the new value smoothly over various time periods. In our eyes, this behavior is a better description of large, complex and dynamic systems such as an economy.\\

After transforming all remaining variables to ensure stationarity and discarding time series with too many NA values, the full data set employed contains 209 time series and covers 234 time periods from 1959Q3 to 2017Q4. This data set is assumed to contain all necessary information to inform the switching mean process of industrial production growth.\\

The factor model outlined in \autoref{sec:dfm} is then applied to this data set. Comparing BICs for models with 1-25 factors points into the direction of a model with seven factors. Thus, the 7-factor model is described in more detail below. Guided by economic theory, a MS model with two states -- interpreted as recessionary and expansionary periods -- is implemented as in previous literature.\footnote{A more sophisticated approach would implement a marginal likelihood based grid search over various combinations of number of factors, number of states and different prior values.}. The estimated factors are then used as exogenous predictors in the MS framework to model the regime switching behavior interpreted as US business cycles.\\

Identification of the MS model is achieved by imposing the identifying restriction $\mu_1 > \mu_2$. Identification of the factor model is achieved by restricting the factor loadings matrix $\Lambda$ to have zeros above the diagonal. In theory, this might lead to problems when the first $r$ variables gain too much influence and weight when employed in estimating the factors. However, a large enough number of time series should prohibit this scenario and estimating 7 factors on more than 200 time series minimizes the risk to a certain degree. In addition, the estimated factor loadings matrix is not extremely sparse, again counteracting these possible threats to identification (see \citealp{kaufmann2019bayesian}).

\begin{figure}[t]
\begin{subfigure}{\textwidth}
  \centering
  \includegraphics[width=0.8\linewidth]{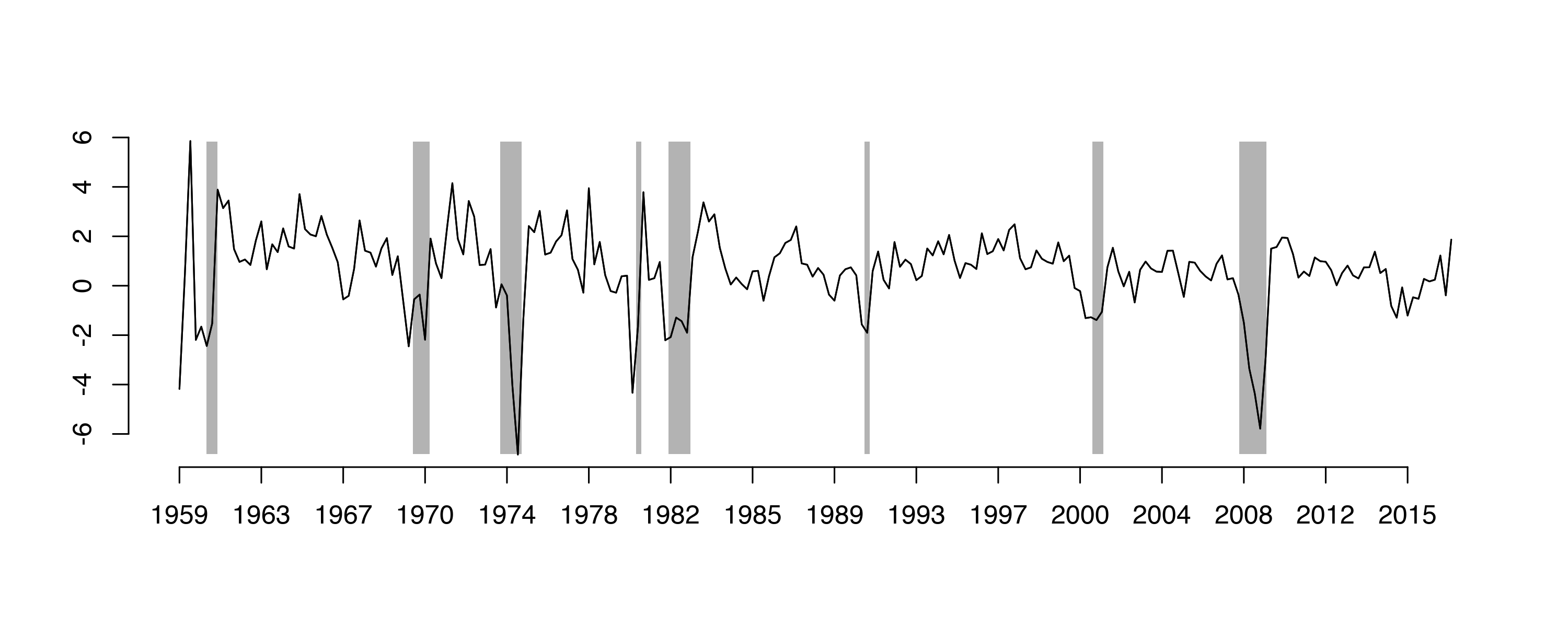}
  \caption{Industrial production growth.}
  \label{fig:indpro}
\end{subfigure}\\
\begin{subfigure}{\textwidth}
  \centering
  \includegraphics[width=0.8\linewidth]{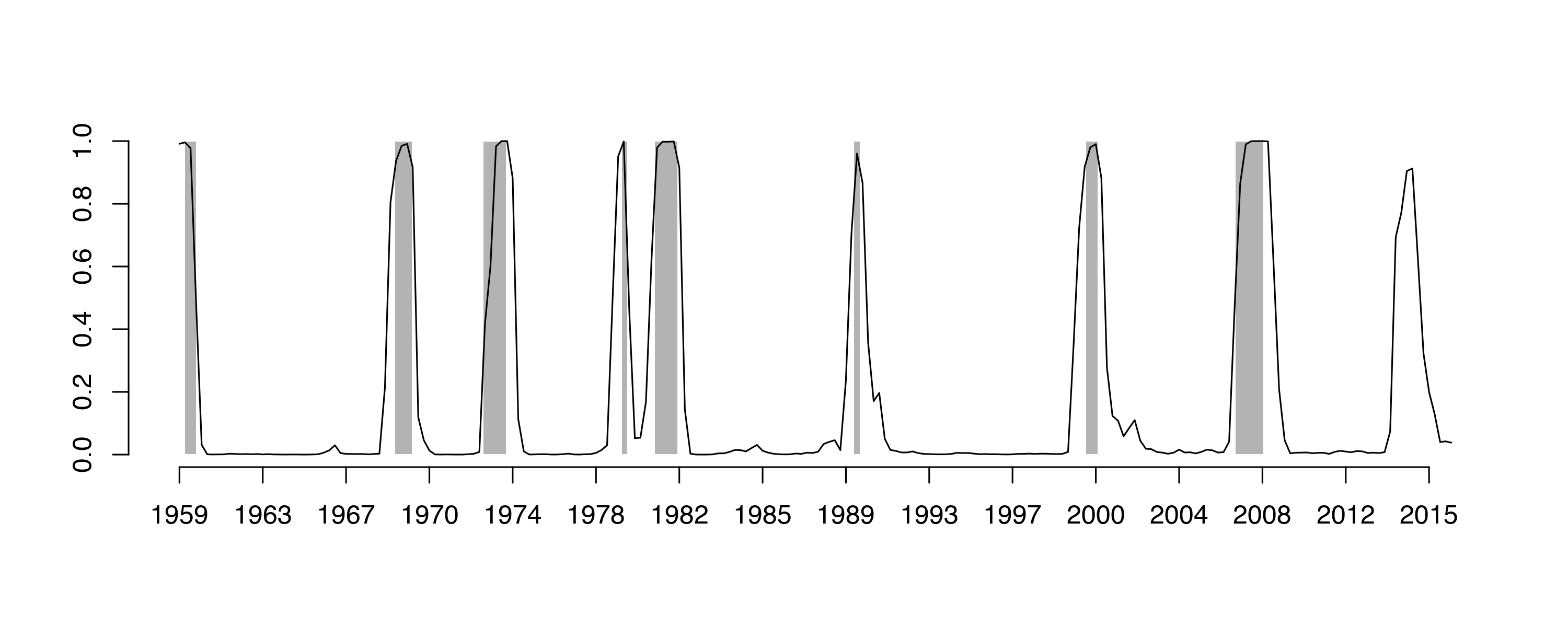}
  \caption{Smoothed state probabilities.}
  \label{fig:filterprobs}
\end{subfigure}
\caption{Input and output from Markov switching model with NBER recession dates.}
\label{fig:indpro_filterprobs}
\end{figure}

\subsection{Results}

The sampling algorithm is iterated $50,000$ times after discarding the first $50,000$ draws as burn-in. Convergence of the MCMC sampler is usually very good. \autoref{fig:infoset} and \autoref{fig:factors} show the full information set used to estimate the factors as well as the factors posterior means used to inform the switching process.\footnote{To be clear, in this exercise the factor means in period $t$ inform the switching process in period $t$. This corresponds to a delay parameter $d=0$. It is possible to look at leading indicators by setting $d>0$. However, this leaves the results qualitatively unchanged in this setup. Thus, we do not show these additional results for brevity reasons.}\\

\begin{table*}[t]
\centering
\begin{threeparttable}
        \caption{AR Process Parameter Estimates}
\label{tab:ar_results}
\setlength\tabcolsep{15pt}
\renewcommand\arraystretch{1.3}
  \begin{tabular}{cd{3.3}d{3.3}d{3.3}}
    \toprule
   & \thead{Est.} &  \multicolumn{2}{c}{\textbf{HPD Region}} \\ 
      \cmidrule(lr{1em}){2-2}      \cmidrule(lr{1em}){3-4}

  $\mu_{\text{Recession}}$ & -0.39 & -0.99 & 0.40 \\  
  $\mu_{\text{Expansion}}$ & 0.70 & 0.26 & 0.97 \\ 
  $\sigma^2$ & 1.40 & 1.14 & 1.66 \\ 
  $\phi_1$ & 0.53 & 0.37 & 0.69 \\ 
  $\phi_2$ & -0.17 & -0.29 & -0.05 \\ 
  $\phi_3$ & 0.06 & -0.05 & 0.18 \\ 
  $\phi_4$ & -0.14 & -0.24 & -0.04 \\ 
\toprule
\end{tabular}
\begin{tablenotes}
\item \small \textit{Note:} Estimates correspond to the posterior medians. 90\% HPD regions are reported.
\end{tablenotes}
\end{threeparttable}
\end{table*}

The estimated factors explain a variance share of 34\% when averaging over time and series. This is comparable to related literature. For instance, \citet{kaufmann2019bayesian} explain around 38\% of the variance of 282 time series using a model with 20 factors. However, the explained variance share varies significantly over time and between series. The interquartile range of the explained variance share over time lies between 23\% and 45\%. The factors explain over 90\% of the variance of 6\% of the time series, over 80\% of 10\% of the series and more than 50\% of the variance of 36\% of the series.\\

Interpreting the factors is more of an art than a science. However, to give a brief intuition on how the factors \textit{could} be interpreted, a table with the time series with the highest absolute factor loadings is provided in \autoref{app:highest}. This table is used to name the factors in the results presented below. However, when interpreting factors, there is usually no clear cut and distinct interpretation possible as we do not restrict time series to only load on one factor. All in all, naming the factors is mainly done to achieve increased ease of presentation, but should be taken with a grain of salt.\\

\autoref{fig:indpro} provides the industrial production growth rate time series used in the analysis. \autoref{fig:filterprobs} plots the estimated posterior median of the transition probabilities. Both plots show recession dates as published by NBER as shaded areas. The FAMS model is able to capture the recessions rather well. Interestingly, the model detects a recession in 2014 that is not classified as a recession by NBER. The corresponding estimates of the parameters of the AR process can be found in \autoref{tab:ar_results}. The intercept estimates point into the direction of one state with rather volatile growth rates of industrial production. A large part of the intercept posterior density of this state also lies in the negative spectrum. Thus, we proceed to label this state \textit{Recession}. On the other hand, the second state can be characterized by consistently positive growth rates and is thus labeled \textit{Expansion}.\\

\begin{figure}
    \centering
    \includegraphics[width=0.45\textwidth]{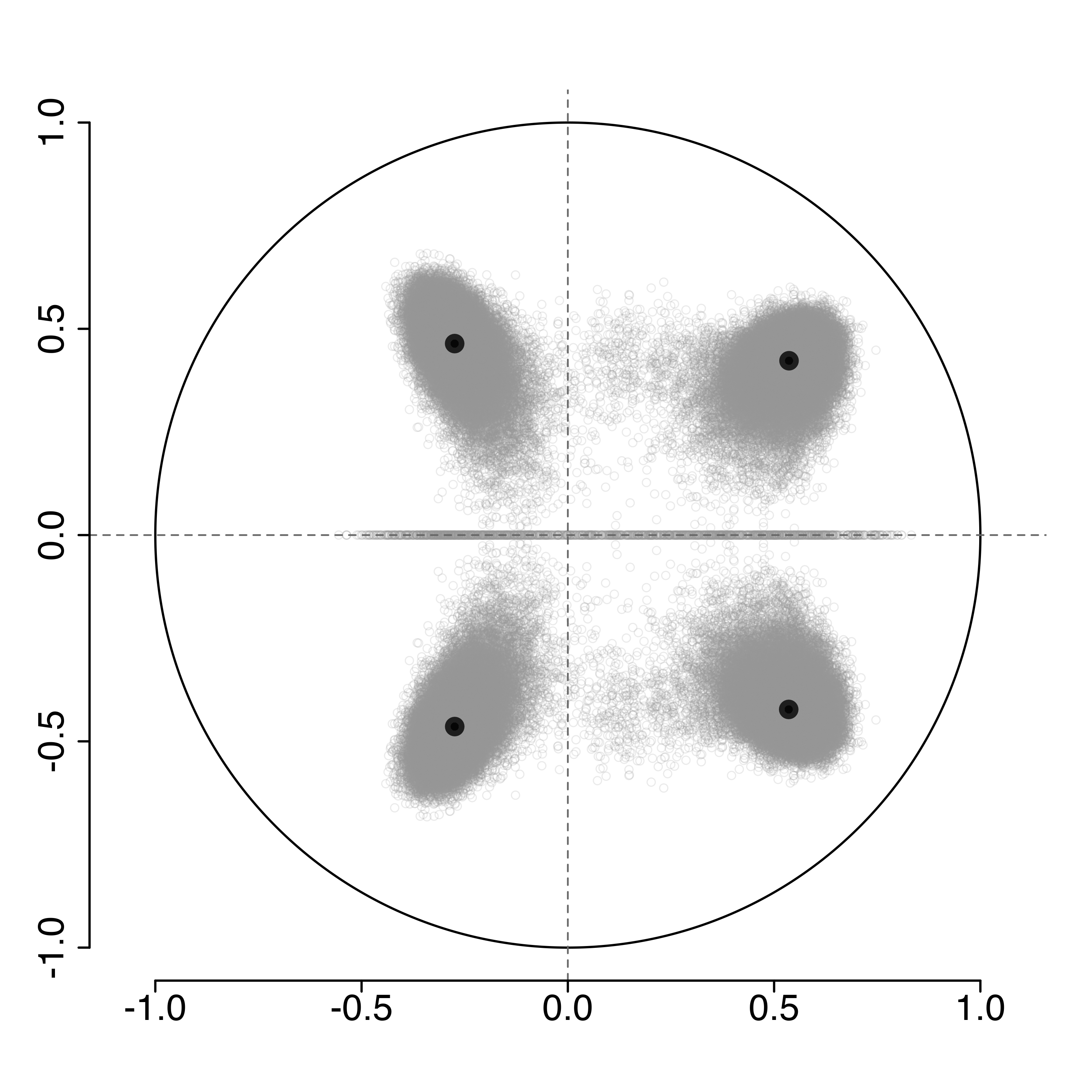}
    \caption{Stationarity plot for AR(4) process.}
    \label{fig:stationary}
\end{figure}

The estimated uncentered AR(4) process seems to be sufficiently stationary. This can be easily inspected visually by rewriting the AR(4) process as AR(1) process. This results in the so called \textit{state space} or \textit{companion} form of the process. If the eigenvalues of the resulting companion matrix lie well inside the unit circle, the process is considered stationary. This exercise is depicted in \autoref{fig:stationary}, where the abscissa denotes the real part and the ordinate the imaginary part of the eigenvalues. Bold black dots denote the median eigenvalues.\\

\begin{figure}
    \hspace*{0.3cm}
    \includegraphics[width=0.7\textwidth]{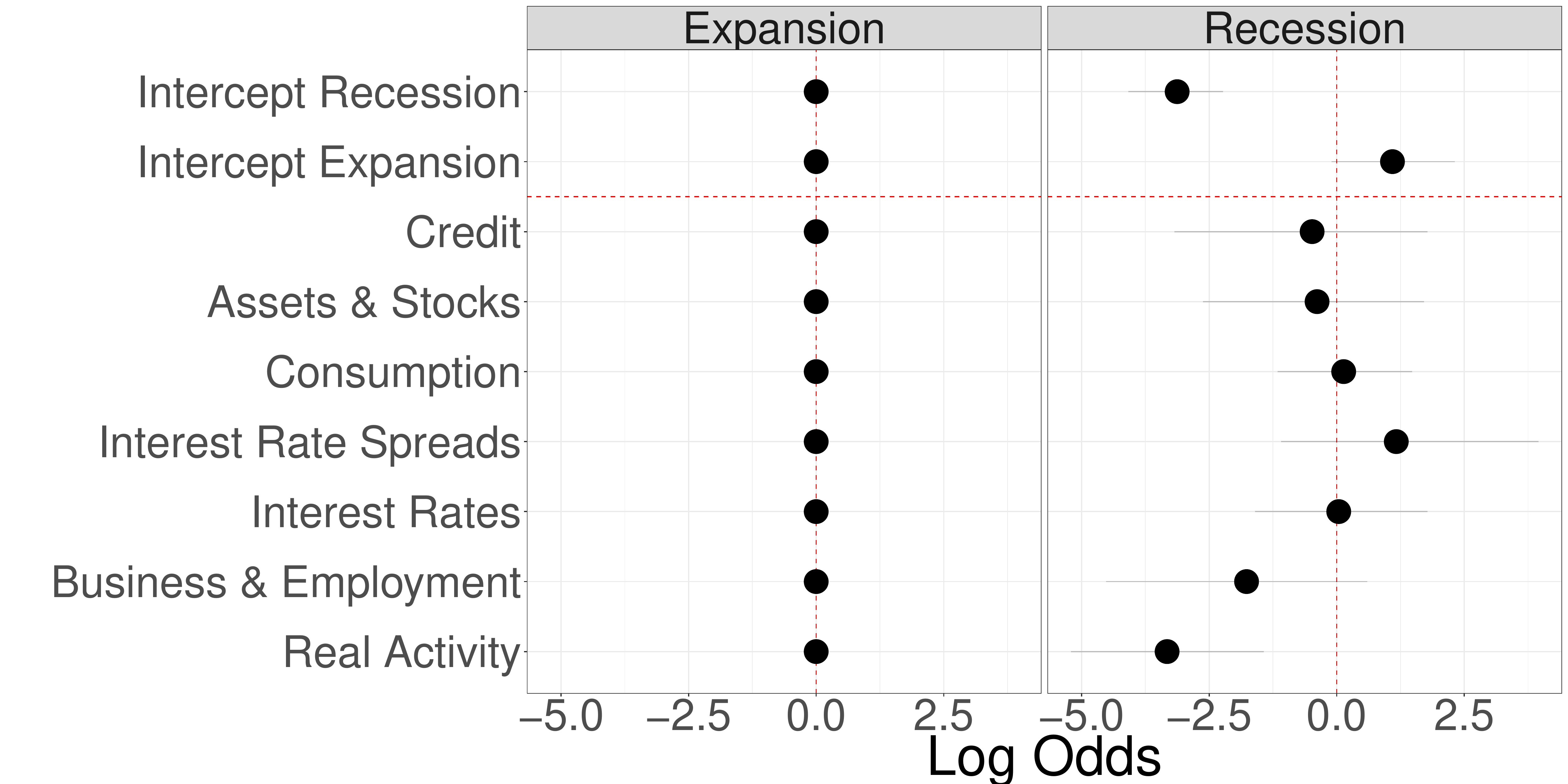}
    \caption{Estimated posterior medians of $\gamma$ and $\beta$. }
    \label{fig:coefs}
\end{figure}

Finally, the results of the multinomial logit part of the FAMS model are provided graphically in \autoref{fig:coefs}.\footnote{Numerical results are available from the authors upon request.} The states are labeled ''Expansion'' and ''Recession'' corresponding to the estimated posterior mean growth rates. ''Expansion'' is the baseline state in this setup, thus all coefficients are set to 0 for identification reasons as discussed earlier. The black dots indicate posterior means and the grey lines show the 90\% HPD region of the posterior density of the MCMC draws. For this estimation run, we set $\omega_{\psi}=0.6$ to apply a medium amount of shrinkage on the factors.\\

A few observations are worth to be mentioned: First, not many of the coefficients are heavily shrunk to zero, indicating a rather large amount of information in the factors that is relevant to inform the switching process in the MS model. Secondly and not surprisingly, the highest correlation of recessions as measured per a real activity indicator is shown by the factors representing real economic activity (\textit{Business \& Employment} and \textit{Real Activity}). Third, \textit{Interest Rate Spreads} seem to have quite some predictive power when it comes to explaining recessions in this data set. This is in line with a large strain of literature connecting real economic recessions and risk premiums on the financial markets (\citealp{gilchrist}; \citealp{salido}).\\

For future applications, it will be interesting to estimate the FAMS with switching variances in stock market or interest rate applications as well as to experiment with various amounts of shrinkage. It is also possible to estimate $\omega_{\psi}$ from the data as demonstrated for instance in \citet{malsiner2016model}.\\

\section{Concluding remarks}
\label{sec:conclusions}

In this article, we develop a factor-augmented Markov switching model with time varying transition probabilities. We assume that a set of latent factors informs the switching process through a multinomial logistic prior setup. This alleviates problems arising from a high-dimensional data set where a large amount of variables might potentially be relevant to model the Markov switching process. Estimation of the factors and the MS model is carried out in a Bayesian framework through Gibbs sampling. In a first step, the FAMS model is applied to artificial data sets. These simulation studies underline the benefits of the FAMS when compared to similar regime switching frameworks. In addition, a real data application highlights the potential of the FAMS model in business cycle research.\\
Future avenues for research include for instance a thorough cross-model comparison of forecasting abilities in different environments and additional applications that test the usefulness of the FAMS framework in the context of switching variances. It will be interesting to see how the model performs in other real world applications where the goal is for instance to model interest rates (\citealp{ang2002regime}) or stock market returns (\citealp{schaller1997regime}).\\
To summarize, we hope the FAMS model will provide a proper modeling framework for fields where Markov switching models are commonly employed in data rich environments. Rather commonly, applications will thus be found in economics and finance. However, the modeling framework might also be useful in fields like medicine, where Markov switching models are used as an early detection system for influenza (\citealp{beneito2008}) or in accident analysis where MS models are used to model vehicle accicdent frequencies (\citealp{malyshkina2009markov}).

\small{\scfont\setstretch{0.85}
\addcontentsline{toc}{section}{References}
\bibliographystyle{custom.bst}
\bibliography{lit}}
\clearpage
\appendix

\setcounter{equation}{0}
\renewcommand\theequation{A.\arabic{equation}}

\section{Auxiliary Mixture Sampling of the MNL coefficients}
\label{sec:mnl_estimation}
This section provides a short overview of the sampling technique employed to simulate from the posterior distribution of the multiomial logit coefficients $\beta$ and $\gamma$. We implement an auxiliary mixture sampler resulting from the partial dRUM representation of the multinomial logistic regression in the style of \citet{fruhwirth2010data} as follows:\\

Let $y_i$ ($i=1,\ldots,N$) be an independent sequence of categorical data where $y_i$ can take one of $m+1$ unordered values. Denote these values or categories by $L = \{0,\ldots,m\}$ and, for any $k$, the set of all categories except $k$ by $L_{-k} = L\setminus\{k\}$. Assume that the observed categorical outcomes $y_i$ result from underlying latent utility process governed by (continous) latent utilities $y^u_{k,i}$. The standard latent variable representation of the multinomial logistic model following \citet{mcfadden1974frontiers} can be written as

\begin{equation}
        y^u_{ki} = x_i\beta_k + \delta_{ki},~~~k=0,\ldots,m\\
\end{equation}

where

\begin{equation}
\label{eq3}
    y_i = k \Leftrightarrow y^u_{ki} = \max_{l \in L} y^u_{li}.
\end{equation}

Thus, the observed category corresponds to the category with the highest latent utility. If the error terms $\delta_{ki}$ follow an extreme value type I distribution, the multinomial logistic regression model results as the marginal distribution of $y_i$. As shown in \citet{fruhwirth2010data}, the latent utilities $y^u_{ki}$ can be sampled simultaneously from

\begin{equation}
    y^u_{ki} = -\text{log}\Big(-\frac{\text{log}(U_i)}{1+\sum_{l=1}^m \lambda_{li}} - \frac{\text{log}V_{ki}}{\lambda_{ki}}I\{y_i \neq k\}\Big),
\end{equation}

where $U_i$ and $V_{1i}, \ldots, V_{mi}$ are $m+1$ independent uniform random numbers in [0,1] and $\lambda_{li} = \text{exp}(x_i\beta_l)$ for $l = 1,\ldots,m$. This corresponds to the standard RUM representation of the multinomial logistic regression.\\
Note that \autoref{eq3} can be rewritten as

\begin{equation}
    y_i = k \Leftrightarrow y^u_{ki} > y^u_{-k,i},~~~y^u_{-k,i} = \max_{l \in L} y^u_{li}.
\end{equation}

Hence, we observe category $k$ if and only if $y_{ki}^u$ is larger than the maximum of all other utilities. This makes it possible to construct another set of latent variables $w_{ki}$ that is defined as the difference between $y^u_{ki}$ and $y^u_{-k,i}$. Note that it directly follows that $y_i = k$ if and only if $w_{ki}>0$. The latent variables $w_{ki}$ thus make it possible to construct binary observations $d_{ki} = I\{y_i = k\}$ whenever $w_{ki} > 0$:

\begin{equation}
    w_{ki} = y^u_{ki} - y^u_{-k,i}, ~~~ d_{ki} = I\{w_{ki}>0\}.
\end{equation}

\citet{fruhwirth2010data} show that the distribution of $w_{ki}$ has an explicit form for the multinomial logistic model and derive the partial dRUM representation of the multinomial logit as

\begin{equation}
    w_{ki} = x_i\beta_k - \text{log}(\lambda_{-k,i}) + \epsilon_{ki}
\end{equation}

where $\epsilon_{ki}$ follows a logistic distribution and

\begin{equation}
    \text{log}(\lambda_{-k,i}) = \sum_{l \in L_{-k}}\lambda_{li}.
\end{equation}

Thus, the problem of sampling $\beta_k$ reduces to sampling regression coefficients from a linear regression with logistic errors. Various sampling methods have been proposed to accomplish this. For instance, \citet{holmes2006bayesian} represent the logistic distribution of $\epsilon_{ki}$ as infinite scale mixture of normals, resulting in a computationally rather demanding sampler. \citet{Scott2011} applies independence MH steps to sample $\beta_k$ using a normal proposal with variance $\frac{\pi^2}{3}$. In the present paper, we use the finite scale mixture approximation of the logistic distribution proposed in \citet{fruhwirth2010data}. With this sampling method, the problem collapses to sampling coefficients from a normal linear regression model with heteroskedastic errors. Hence, the resulting Gibbs sampler can be implemented in a computationally very efficient way. Prior distributions and posterior simulation are discussed in \autoref{sec:prior} and \autoref{sec:post}.

\section{Posterior Simulation of $\beta$}
\label{ngprior}
The posterior distributions of $\lambda_j^2$ and $\psi_{i,h}$ are of well-known form and can be derived as

\begin{equation}
    \begin{split}
        \pi(\lambda^2_{\psi,h}~|~\cdot) &\sim G(g_1,d_h)\\
        \pi(\psi_{i,h}~|~\cdot) &\sim GIG(\omega_\psi-0.5,~ \beta_{i,h}^2,~ \lambda^2_{\psi,h} \omega_\psi)\\
        g_1 &= \omega_\psi r + c_0\\
        d_h &= c_1 + \frac{\omega_\psi}{2\sum_{j=1}^m \psi_{i,h}},
    \end{split}
\end{equation}

where $r$ is the number of factors entering the model and $GIG$ denotes the Generalized Inverse Gaussian distribution. The posteriors of $\tau_{i,j}$ and $\lambda^2_{\tau,i}$ can be derived in a similar fashion. \citet{hormann2014generating} provide an efficient adaptive rejection sampling algorithm that makes it possible to easily draw from the $GIG$. We use the R package \textit{GIGrvg} (\citealp{leydold2015gigrvg}) in our computations to employ this algorithm. 

\clearpage
\section{Highest loading time series per factor}
\label{app:highest}
\begin{table}[ht]
\centering
\footnotesize
\begin{tabular}{l}
  \\ 
     \hline
  Industrial Production:  Manufacturing (SIC) (Index 2012=100) \\ 
  Nonfarm Business Sector:  Real Output \\ 
  Business Sector:  Real Output \\ 
  Real Gross Domestic Product \\ 
  Real private fixed investment  \\ 
  IP:Final products Industrial Production: Final Products (Market Group) (Index 2012=100) \\ 
  IP:Consumer goods Industrial Production: Consumer Goods (Index 2012=100) \\ 
  Durable Consumer Goods (Index 2012=100) \\ 
  Unemployment Rate less than 27 weeks (Percent) \\ 
   \hline
Total Business Inventories (Millions of Dollars) \\ 
  All Employees:  Wholesale Trade (Thousands of Persons) \\ 
  Capacity Utilization:  Manufacturing (SIC) (Percent of Capacity) \\ 
  All Employees:  Service-Providing Industries (Thousands of Persons) \\ 
  Shares of gross domestic product:  Gross private domestic investment: Change
in private inventories \\ 
  All Employees:  Trade, Transportation \& Utilities (Thousands of Persons) \\ 
   Emp:Nonfarm All Employees: Total nonfarm (Thousands of Persons) \\ 
   All Employees: Total Private Industries (Thousands of Persons) \\ 
  Moody’s Seasoned Aaa Corporate Bond Minus Federal Funds Rate \\ 
  All Employees:  Durable goods (Thousands of Persons) \\ 
   \hline
6-Month Treasury Bill: Secondary Market Rate (Percent) \\ 
  1-Year Treasury Constant Maturity Rate (Percent) \\ 
  3-Month Treasury Bill: Secondary Market Rate (Percent) \\ 
  3-Month AA Financial Commercial Paper Rate \\ 
  5-Year Treasury Constant Maturity Rate \\ 
  Effective Federal Funds Rate (Percent) \\ 
  10-Year Treasury Constant Maturity Rate (Percent) \\ 
  Moody’s Seasoned Aaa Corporate Bond Yield (Percent) \\ 
  Moody’s Seasoned Baa Corporate Bond Yield (Percent) \\ 
  6-Month Treasury Bill Minus 3-Month Treasury Bill, secondary market (Percent) \\ 
   \hline
Moody’s Seasoned Aaa Corporate Bond Minus Federal Funds Rate \\ 
  All Employees:  Education \& Health Services (Thousands of Persons) \\ 
  All Employees:  Government:  State Government (Thousands of Persons) \\ 
  All Employees:  Government:  Local Government (Thousands of Persons) \\ 
  3-Month Commercial Paper Minus 3-Month Treasury Bill, secondary market \\ 
  All Employees:  Other Services (Thousands of Persons) \\ 
  1-Year Treasury Constant Maturity Minus 3-Month Treasury Bill, secondary market \\ 
  All Employees:  Government (Thousands of Persons) \\ 
  Average Weekly Hours of Production and Nonsupervisory Employees:  Manufacturing \\ 
  Capacity Utilization:  Manufacturing (SIC) (Percent of Capacity) \\ 
   \hline
\end{tabular}
\caption{Ten Time Series with Highest Factor Loadings Factors 1-4} 
\end{table}  
   
\begin{table}[ht]
\centering
\footnotesize
\begin{tabular}{l}
   \hline
Personal consumption expenditures:  Goods  \\ 
  Personal consumption expenditures:  Nondurable goods \\ 
  Consumer Price Index for All Urban Consumers:  Commodities \\ 
   \\ 
  Consumer Price Index for All Urban Consumers:  Transportation \\ 
  Consumer Price Index for All Urban Consumers:  All Items Less Food \\ 
  Consumer Price Index for All Urban Consumers:  All Items \\ 
   \\ 
  Personal Consumption Expenditures: Chain-type Price Index  \\ 
  Personal consumption expenditures:  Nondurable goods:  Gasoline and other energy goods \\ 
   \hline
S\&P’s Common Stock Price Index:  Industrials \\ 
  S\&P’s Common Stock Price Index:  Composite \\ 
  S\&P’s Composite Common Stock:  Dividend Yield \\ 
  Real Total Financial Assets of Households and Nonprofit Organizations \\ 
  Real Assets of Households and Nonprofit Organizations excluding Real Estate Assets \\ 
  S\&P’s Composite Common Stock:  Price-Earnings Ratio \\ 
   \\ 
  Real Total Assets of Households and Nonprofit Organizations \\ 
  Nikkei Stock Average \\ 
  Moody’s Seasoned Aaa Corporate Bond Minus Federal Funds Rate \\ 
   \hline
Nonrevolving consumer credit to Personal Income \\ 
  Industrial Production:  Manufacturing (SIC) (Index 2012=100) \\ 
  Business Equipment (Index 2012=100) \\ 
  Durable Materials (Index 2012=100) \\ 
  IP:Final products Industrial Production: Final Products (Market Group) (Index 2012=100) \\ 
  Total Real Nonrevolving Credit Owned and Securitized, Outstanding \\ 
  Materials (Index 2012=100) \\ 
  Total Consumer Credit Outstanding \\ 
  Shares of gross domestic product:  Exports of goods and services (Percent) \\ 
  All Employees:  Financial Activities (Thousands of Persons) \\ 
   \hline
\end{tabular}
\caption{Ten Time Series with Highest Factor Loadings Factors 5-7} 
\end{table}

\end{document}